\documentstyle[11pt,newpasp-1,twoside,epsf,psfig]{article}
\begin{document}
\begin{center}
\vspace{-3cm}
\title{RECONSTRUCTION OF OBJECTS BY DIRECT DEMODULATION}
 \author{TI-PEI LI and MEI WU}
\affil{ High Energy Astrophysics Laboratory, Institute of High Energy Physics\\
Chinese Academy of Sciences, Beijing, China}
\end{center}
\begin{abstract}
High resolution reconstruction of complicated objects from incomplete and
noisy data can be achieved by solving modulation equations iteratively under 
physical constraints. This direct demodulation method is a powerful technique 
for dealing with inverse problem in general case. Spectral and image 
restorations and computerized tomography are only particular cases of 
general demodulation. It is possible to reconstruct an object in higher 
dimensional space from observations by a simple lower dimensional instrument
through direct demodulation. Our simulations show that wide field and 
high resolution images of space hard X-rays and soft $\gamma$-rays can be 
obtained by a collimated non-position-sensitive detector without coded 
aperture masks.
\end{abstract}

\section{Introduction}
An observation to an object in space \mbox{\boldmath $x$} with 
an intensity distribution $f(\mbox{\boldmath $x$})$ can be seen as a modulation process
of signal from object by observation instrument,
the observed data $d(\mbox{\boldmath $\omega$})$ is an output after 
the modulation, where \mbox{\boldmath $\omega$} denotes the parameters 
determining the state of observation. The modulation equation 
relating the data to object can be written as
\begin{equation} 
\int p(\mbox{\boldmath $\omega,x$})f(\mbox{\boldmath $x$})d\mbox{\boldmath $x$}
 = d(\mbox{\boldmath $\omega$})~,
\end{equation}
where a value $p(\mbox{\boldmath $\omega ,x$})$ of  the integral kernel 
(modulation function) is the modulation coefficient or response coefficient 
of the instrument to a point 
\mbox{\boldmath $x$} of the object space during an observation 
\mbox{\boldmath $\omega$}.

  Generally, the data space \mbox{\boldmath $\omega$} is not the same kind 
of the object space \mbox{\boldmath $x$}. For an imaging formation system 
consisting of position-sensitive detectors, the data space 
\mbox{\boldmath $\omega =x'$} is a two-dimensional space correspondent to 
the object space \mbox{\boldmath $x$} and the observed data 
$d(\mbox{\boldmath $x'$})$ is
an image of the object $f(\mbox{\boldmath $x$})$, if the modulation function 
(point-spread function) is space-invariant, 
$p(\mbox{\boldmath $x',x$})=p(\mbox{\boldmath $x'-x$})$, 
then the modulation equation (image formation equation) can be written 
as the following convolution formula
\begin{equation} 
\int p(\mbox{\boldmath $x'-x$})f(\mbox{\boldmath $x$})d\mbox{\boldmath $x$} 
\equiv p*f 
= d(\mbox{\boldmath $x'$})~. \end{equation}
The image formation Equation (2) is a particular case of the general 
modulation Equation (1), and the image restoration (restoring the object 
$f(\mbox{\boldmath $x$})$ from the observed image $d(\mbox{\boldmath $x'$})$), 
then, is a particular case of object reconstruction.

  Dividing object space into $N$ bins, for $M$ observed values 
$d(k)$, $k=1,...,M$, the modulation Equation (1) or image 
formation Equation (2) in discretization form constitute an algebraic 
equation system
\begin{equation} \sum_{i=1}^{N} p(k,i)f(i) = d(k) \hspace{5mm}    (k=1,...,M).
\end{equation}

  In this paper we introduce the direct demodulation method of 
reconstructing object \mbox{\boldmath $f$} from observed data 
\mbox{\boldmath $d$}. 
This method can be applied to any kind of observation described 
by Equation (3). For a start we will give a brief summation of various 
reconstruction methods in classification.

\section{Methods of Reconstruction}

\subsection{Linear Transformation}
For observations by an image formation system, taking Fourier 
transforms of both sides of the image formation Equations (2) and
taking advantages of the convolution theory of the Fourier transform, we 
have ${\cal F}p\cdot{\cal F}f = {\cal F}d$, then with the inverse 
transform the object can be derived as $f = {\cal F}^{-1}({\cal F}d/{\cal F}p)$.

  In the case of observed values \mbox{\boldmath $d$} being projections, i.e. 
line integrals of object $f(\mbox{\boldmath $x$})$ along lines 
$L: a = \mbox{\boldmath $\xi$} \cdot \mbox{\boldmath $x$}$  (see Figure 1)
\[ \int_{L_{a,\phi}} f(\mbox{\boldmath $x$}) d\mbox{\boldmath $x$} = 
\int_{L_{a, \phi}} 
f(\mbox{\boldmath $x$})
 \delta (a-\mbox{\boldmath $\xi$}\cdot \mbox{\boldmath $x$}) 
d\mbox{\boldmath $x$} = d(a,\phi) \] 
then \mbox{\boldmath $d$} is a sample of the Radon transform (Radon 1917), 
$d = {\cal R}f$, with the inverse 
transform we can reconstruct the object from projections, $f = {\cal R}^{-1}d$.
The Radon transform is the mathematical basis of the computerized 
tomography (CT) technique.
\begin{figure}
\vspace{1cm}
\hspace{5cm}
\psfig{figure=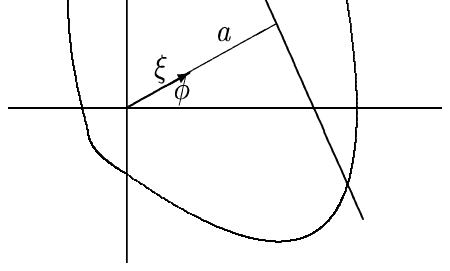,width=20cm,angle=0
}
\vspace{-24.5cm}
\caption{Projection along line $L$ in two-dimensional Radon transform.}
\label{picture}
\end{figure} 

  The cross-correlation function 
\begin{equation}
 c(\mbox{\boldmath $x$})=\int p(\mbox{\boldmath $\omega,x$})
d(\mbox{\boldmath $\omega$})d\mbox{\boldmath $\omega$} \end{equation} 
is a distribution
in the object space. In general case we can use the correlation
calculation to perform transformation from data space to object space.
Correspondent to the modulation Equation (3), the discrete correlation transform
is
\begin{equation} c(i) = \sum_{k} p(k,i)d(k)~. \end{equation} 
There exists another form of correlation transform defined as 
\begin{eqnarray} c^{*}(i) & = & \frac{A}{M} \sum_{k=1}^{M} (p(k,i)-\bar{p})
 (d(k)-\bar{d}) \nonumber \\ & = & A' [M \sum_{k} p(k,i)d(k) - 
 \sum_{k} p(k,i) \sum_{k} d(k)]~. \end{eqnarray} 
If the object is an isolated point source with intensity $f_{s}$ at 
point $i_{s}$, then the correlation function $c^{*}(i) = f_{s} \cdot c_{s}
 (i;i_{s})$,
where $c_{s}(i;i_{s})$ is the resolution function for point $i_{s}$
\begin{equation} c_{s}(i;i_{s}) = A' [M \sum_{k} p(k,i) p(k,i_{s}) -
 \sum_{k} p(k,i) \sum_{k} p(k,i_{s})]~. \end{equation}
The distribution of $c^{*}(i)$ will appear a bulge around the bin $i_{s}$,
provided that the observations get necessary information of object (the
observations are properly arranged and the number $M$ of observations
is large enough). The normalization constant can be chosen as
$A' = 1/[M \sum_{k} (p(k,i))^{2} - ( \sum_{k} p(k,i))^{2}]$,
then the maximum value of the correlation
function, $c^{*}(i_{s})$, equals the source strength, $f_{s}$. The width of the
resolution function $c_{s}(i)$ represents the resolution ability of the 
observations, which can be defined as the intrinsic resolution of the
observations.

\subsection{Statistical Reconstruction}
Another important category of reconstruction methods is to use 
a certain statistic (e.g. the maximum likelihood ratio (Pollock et al. 1981))
instead of the object itself to represent main characters of object by some 
degree or to use values of a statistic (e.g. the $\chi^{2}$ quantity)
instead of the observation modulation equations in selecting possible
object distributions. If the object distribution can be described 
by a theoretical or empirical expression $f(i) = g(i;\mbox{\boldmath $a$})$, 
where $\mbox{\boldmath $a$}=(a_{1},...,a_{l})$ are $l$ undetermined parameters,
the least-squares reconstruction $g(i;\mbox{\boldmath $a$})$  $(i=1,...,N)$ 
satisfies the condition 
\[ \chi^{2}(\mbox{\boldmath $a$})=\sum_{k}\{[d(k)-\sum_{i} p(k,i) 
g(i;\mbox{\boldmath $a$})]/\sigma_{k}\}^{2} = \min \]
with $\sigma_{k}$ being 
the statistical error of $d(k)$. The maximum-entropy method (for example,
see Gull \& Daniell 1978) is to choose a solution $f(i)~ (i=1,...,N)$ from all 
satisfying the statistical criterion 
\[ \chi^{2}(\mbox{\boldmath $f$})=\sum_{k=1}^{M} \{[\sum_{i} p(k,i)f(i)-d(k)]/
\sigma_{k}\}^{2}=M \]
by the condition that the information entropy 
\[  S(\mbox{\boldmath $f$})=-\sum_{i} f_{i}\log f_{i} = \max. \]

\subsection*{2.3. Direct Demodulation}
The most straightforward way to evaluate an object 
\mbox{\boldmath $f$} 
from observed data \mbox{\boldmath $d$} should be directly solving the 
modulation equation connecting
\mbox{\boldmath $d$} with \mbox{\boldmath $f$}, i.e. direct demodulation, 
which we will discuss in this paper.

  Using a statistic instead of original modulation equations to make     
reconstruction will cause loss of information containing in the 
equations about object and observations, then that of sensitivity and 
resolution ability of reconstruction. The least-squares method is
model dependent, not applicable in general case. Reconstruction by
Fourier or Radon transform is mathematically based on modulation 
equation, but only can work on a special kind of observation,
i.e. image formation or projection. In application of Fourier or Radon
transform to incomplete and noisy data in practice, in particular  
to that with poor statistics and low ratio of signal to noise,  
in order to derive a reasonable reconstruction  
a certain treatment (e.g. frequency space filtering) should be
applied, causing loss of information. Correlation calculation can perform 
transformation from data space to object space in general case,
thus being a useful tool in dealing with reconstruction problems;
but the correlation function $c(i)$ or $c^{*}(i)$ is not a good representation
of an object, there may exist regions with negative $c^{*}(i)$
even if the object is defined only in a positive space.

  Before introducing the direct method of reconstruction, we will
firstly compare various methods of solving a linear equation system.

\section{Solutions of Linear Algebraic Equations}
The modulation equation system (3) can be rewritten in matrix
form as
\begin{equation}   \mbox{\boldmath $Pf = d$}~, \end{equation}
where modulation matrix $\mbox{\boldmath $P$}=\{{p(k,i)}\}$, object vector 
$\mbox{\boldmath $f$}=\{f(i)\}$,
data vector $\mbox{\boldmath $d$}=\{d(k)\}, i=1,...,N, k=1,...,M.$

\subsection{Mathematical Solution}
If the modulation matrix \mbox{\boldmath $P$} 
is non-singular and $M=N$, the
exact solution of Eq.(8), $\mbox{\boldmath $f=P$}^{-1}\mbox{\boldmath $d$}$, 
can be found by Gaussian elimination or triangular resolution. 
Iterative method can also be 
used to solve Eq.(8). An iteration algorithm in common use is the 
Gauss-Seidel method with successive relaxation. The approximate solution 
for $l$-th iteration can be calculated by one of the following two formulas
\begin{eqnarray} f_{i}^{(l)} &=& \alpha(d_{i}-\sum_{j=1}^{i-1}p_{ij}
f_{j}^{(l)}-\sum_{j=i+1}^{N}p_{ij}f_{j}^{(l-1)}+(1-p_{ii})f_{i}^{(l-1)}) 
+(1-\alpha )f_{i}^{(l-1)}      \\
f_{i}^{(l)} &=& \frac{\alpha}{p_{ii}}(d_{i}-\sum_{j=1}^{i-1}
p_{ij}f_{j}^{(l)}-\sum_{j=i+1}^{N}p_{ij}f_{j}^{(l-1)})+(1-\alpha)f_{i}^{(l-1)}
  \end{eqnarray}
where the relaxation factor  $0<\alpha\leq1$.
It can be proved that the Gauss-Seidel iteration converges to the exact
solution if the matrix \mbox{\boldmath $P$} is symmetry and positive definite.

   The mathematical solution of a modulation equation system satisfy
the equations accurately. But there always exist errors in modulation 
equations (measurement errors, statistical fluctuation and noises
in data and errors in determining modulation coefficients).
A modulation equation system in a practical problem is usually a
large set of linear equations, its mathematical solutions will
seriously deviate from the true object distribution due to errors
in the equations. A one-dimensional example is displayed
in Figure 2: (a) is an object distribution \mbox{\boldmath $f$}, a line at $i_{0}$
on a uniform background; (b) the ideal data $d_{0}(k)=\sum_{i} p(k,i)f(i)$,
where the triangular distribution presents the shape of the 
modulation function $p(k,i_{0})$; (c) the simulated observed data 
\mbox{\boldmath $d$},
a Monte Carlo sample from (b); (d) and (e) are the correlation 
distribution $c(i)$ and $c^{*}(i)$ respectively; (f) presents the 
mathematical solution of the modulation equations 
\mbox{\boldmath $Pf=d$}
by Gauss-Seidel iterations. From Figure 2f we can see that the
mathematical solution of the modulation equations 
with statistical errors is in an ill-conditioning feature with
violent oscillation, including a lot of nonphysical values of $f<0$.
\begin{figure}
\hspace{3.0cm}
\vspace{-3.0cm}
\psfig{figure=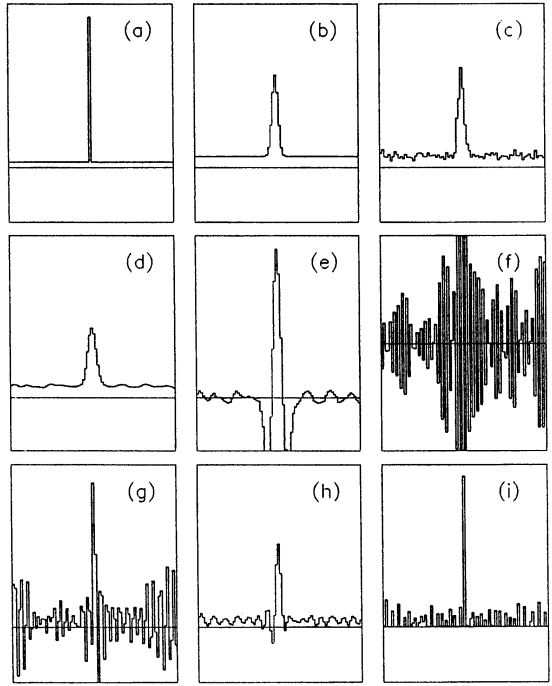,width=10cm,angle=0
}
\vspace{-2cm}
\caption{ Solutions of modulation equations. (a) Object spectrum. 
(b) Theoretical observed spectrum. (c) Simulated observation data.
(d) Correlation distribution $c(i)$. (e) Correlation distribution $c^{*}(i)$.
(f) Mathematical solution of modulation equations. (g) Least-squares 
solution of modulation equations. (h) Solution of correlated modulation
equations of $L=2$. (i) Iterative solution with nonnegative constraint.
All horizontal axes are in the same scale, the horizontal line in 
each plot indicates the zero level, the vertical axes are bounded from
$-200$ to 600 for (a), (e), (g), (h), (i), that from $-100$ to 300 
for (b), (c), (d) and from $-10000$ to 10000 for (f). }
\label{picture}
\end{figure}

\subsection{Statistical Solution}  
If the number of observed values is greater than that of 
the undetermined parameters, $M>N$, the system (3) is 
overdetermined, we can use the least-squares criterion 
$\sum_{k} [\sum_{i}p(k,i)f(i)-d(k)]^{2} =$ min to derive the normal equations as
\begin{equation}  
\sum_{i=1}^{N} p_{1}(i',i)f(i) = c_{1}(i') \hspace{5mm}  (i'=1,...,N), 
\end{equation}
where $p_{1}(i',i)=\sum_{k}p(k,i')p(k,i)$, the right side of the normal 
equation is the correlation function 
$c_{1}(i')\equiv c(i')=\sum_{k} p(k,i')d(k)$.
The mathematical solution of the normal Equations (11), i.e. the
least-squares solution of the modulation Equations (3), can be
taken as an estimation of object in a sense of least-squares
(in a statistical sense).

  The normal equations are just the correlation transform of the 
original modulation Equations (3), we can call them correlated 
modulation equations. For the case of $M\leq N$ we can also  
derive equations of the same form by means of correlation transformation. The 
correlation transformation can be repeatedly performed (Li \& Wu 1992, 1993).
Starting from the original modulation eqs. (3), after $L$-fold   
correlation transforms we can obtain the $L$-order correlated
modulation equations as
\begin{equation} \sum_{i=1}^{N} p_{L}(i',i) f(i) = c_{L}(i') \hspace{5mm}
(i'=1,...,N), \end{equation}
where $p_{L}(i',i)=\sum_{j}p_{L-1}(j,i')p_{L-1}(j,i), 
c_{L}(i')=\sum_{j}p_{L-1}(j,i')c_{L-1}(j)$.
  
  Figure 2g and 2h show the solutions of correlated modulation equations
of $L=1$ and $L=2$, respectively (statistical solutions of modulation
equations).

\subsection{Physical Solution (Constrained Iterative Solution)} 
Setting reasonable physical constraints in the iterative process of 
solving a modulation equation system (3) or correlated modulation equation 
system (12) will help us obtain an estimation of the object distribution 
satisfying some necessary physical conditions (Li \& Wu 1992,1993). 
Corresponding to a certain object,
we can set up a lower bounds \mbox{\boldmath $b$} and upper bounds 
\mbox{\boldmath $u$}, 
for any approximate solution $f^{(l)}(i)$,
\begin{eqnarray} if \hspace{2mm} f^{(l)}(i)&<&b(i), 
             \hspace{3mm} let \hspace{2mm} f^{(l)}(i)=b(i), \nonumber \\
                 if \hspace{2mm} f^{(l)}(i)&>&u(i), 
             \hspace{3mm} let \hspace{2mm} f^{(l)}(i) = u(i). \end{eqnarray}
The final solution should be normalized by the condition 
 $\sum_{k}\sum_{i}p(k,i)f(i)=\sum_{k}d(k)$.
For many objects, nonnegativity is a necessary constraint, i.e.   
$b(i)=0, (i=1,...,N)$. Figure 2i is a constrained iterative solution
for data (c) under the nonnegative condition after $10^{2}$ iterates. 
The constrained iteration process has a good property of convergence and
is not sensitive to small errors in the modulation function. We found
the approximate solution after $10^{4}$ or more iterates or that starting from 
completely different initial values had not any recognized difference 
with Figure 2i. When the modulation function used in iteration being widened 
in a direction and narrowed in the other direction by $20\%$ separately, 
no obvious change was found in the constrained iterative solution,
and when the whole modulation function being widened by $20\%$ 
also almost the same figure of solution was obtained but just the peak height 
increased a little. 

  Exerting the constraints of lower and upper bounds is a kind of nonlinear
control in the iterative precess, proper nonlinear control can depress
the influence of noise, fluctuation and other errors in data effectively 
and strengthen convergence of the iterative process, helping
to derive a satisfactory reconstruction in a physical sense. There exist various
iterative algorithms applicable to seeking physical solutions.
The modulation matrix $p_{L}$ of the correlated modulation Equations (12) is 
symmetry and positive definite, satisfying convergence
condition of the successive relaxation Gauss-Seidel iteration.
Another useful algorithm is the 
Richardson-Lucy iteration (Richardson 1972; Lucy 1974) based on the Bayes theory
\begin{equation} f^{(l)}(i) = f^{(l-1)}(i) \sum_{k} \frac{p(k,i) d(k)}{\sum_{i'} p(k,i')
 f^{(l-1)}(i')}/\sum_{k}p(k,i)~. \end{equation}

  Solutions from probability iterations satisfy the nonnegative condition
and normalization condition of total flux automatically. 
However, only nonnegative constraint is not enough to
control the iterative process to produce a satisfactory reconstruction 
from many observations, in particular from that of poor statistics and low
ratio of signal to noise for complicated objects. In the case of 
object space including negative area, the simple nonnegative constraint 
can not be used at all. 

  Relaxation functions to modify the correction term
gradually as iteration proceeds were suggested for spectral deconvolution
(for example, see Jansson 1984). In oder to fulfil a need for setting 
various kind of physical constraints necessary in practice
we proposed to set boundary constraints directly on each approximate solution 
as shown by expression (13) and found that a direct demodulation 
method in the general case can be constituted by this technique, which we will
introduce in the following section. 

\section{Direct Demodulation Method}
It is obvious that 
in solving modulation Equations (3) or correlated modulation Equations (12)
taking the intensities $\mbox{\boldmath $f$}_{b}$ 
of the diffuse background in the object space as lower bounds, i.e.
\begin{equation}   \mbox{\boldmath $b = f_{b}$} \end{equation}
can make more precise control than using simple nonnegative constraint, 
giving a better estimation of the object.

  In the case of no priori knowledge of the diffuse background, it is 
needed to estimate it from observed data. First we use a successive
procedure similar to CLEAN (H\"{o}gbom 1974) to subtract all contribution of
apparent discrete sources with intensities greater than $f_{m}$ from
the observed data \mbox{\boldmath $d$} and derive background data 
\mbox{\boldmath $d_{b}$}, where
$f_{m}$ can be taken as the minimum intensity of discrete sources 
to be reconstructed. Solving the modulation equations for 
\mbox{\boldmath $d_{b}$},
\mbox{\boldmath $Pf_{b}=d_{b}$}, by an iterative method we finally get 
\mbox{\boldmath $f_{b}$}. In iterations we can 
simply smooth each approximate solution \mbox{\boldmath $f_{b}^{(l)}$} or use  
some kind of continual condition like  
\begin{equation} u(i) = f_{b}^{(l)}(i-1) + \delta, \hspace{3mm}  
b(i) = f_{b}^{(l)}(i-1) - \delta \end{equation}
to depress sudden change of intensity at any point $i$ from its close
neighbour one $i-1$ in the object space, where $\delta=f_{m}/ n_{b}$ 
with $n_{b}$ being the linear size of minimum diffuse structure reconstructed 
by number of bins.

  Subtracting contribution of apparent discrete sources from data can be 
performed with the aid of cross-correlation technique. Find out the 
maximum point $i_{s}$ of the correlation map $c^{*}(i)$ $(i=1,...,N)$ for 
data \mbox{\boldmath $d$} and the intensity excess $f_{s}$ above the background
level at $i_{s}$, remove the contribution of $f_{s}$ from the data by 
$d_{b}(k)=d(k)-p(k,i_{s})f_{s}$, $(k=1,...,M)$. The procedure is repeated for
\mbox{\boldmath $d_{b}$} till $f_{s} < f_{m}$. The excess $f_{s}$ can be 
determined by 
the condition $\sum_{i'}(c_{b}(i')-[c^{*}(i') - f_{s} c_{s}(i',i_{s})])^{2}
=min$, where the sum for $i'$ is accumulated over
a region around $i_{s}$ within the width of the resolution function, 
$c_{b}(i')$ is the extrapolated value of correlated intensities $c^{*}(i)$
outside the region to the point $i'$.

  An example of restorations of a line source superimposed on an
inclined background is shown in Figure 3: (a) is the object distribution,
 (b) the simulated data,
(c) Richardson-Lusy iterative restoration (i.e. a constrained iterative 
solution of modulation equations with only the nonnegative condition), 
(d) restoration by the direct demodulation for correlated modulation 
equations of $L=1$.
\begin{figure}
\hspace{4cm}
\psfig{figure=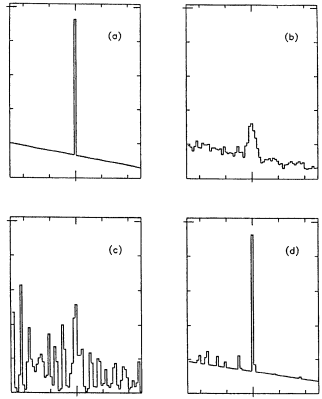,width=7cm,angle=0
}
\vspace{-4.5cm}
\caption{Direct demodulation. (a) Object spectrum. (b) Simulated data.
(c) Richardson-Lusy restoration. (d) Direct restoration.}
\label{picture}
\end{figure}

  For many reconstruction problems in practice, performing 
iterations in two steps with the constraint condition (16) and (15) separately
by the Gauss-Seidel formula with successive relaxation 
to the correlated modulation Equations (12) of $L=1$, or by
the Richardson-Lucy formula to the original modulation
Equations (3), one can get satisfactory reconstructions.
In the case of data having good statistics, solving the original modulation 
Equations (3) by constrained Gauss-Seidel iterations will give high 
resolution reconstruction; but in the other hand, for observations with
severe noise and fluctuation and bad intrinsic resolution,  
higher order correlated modulation Equations (12) can serve to get better 
reconstruction. In iterations for reconstructing diffuse background correlated 
modulation equations of higher order are also recommended.  

\section{Modulation Imaging}
The direct demodulation method can be applied to dealing with
various kind of reconstruction problems, i.e. to solving the inverse problems
of the general modulation Equations (3), where the data space $ $ determined by
apparatus and states of observation can be completely different with
the observed object space \mbox{\boldmath $x$}, and the number of observations, 
$M$, is not necessary to be equal to or greater than that of undetermined 
parameters, $N$ (Li et al. 1993). A general demodulation technique make it possible to design
an experiment by simple instrument to realize the same objective as more 
complicated one does, for example, to obtain a two-dimensional image
by a one-dimensional non-position-sensitive detector ( non-imaging 
system).
\begin{figure}
\hspace{2cm}
\psfig{figure=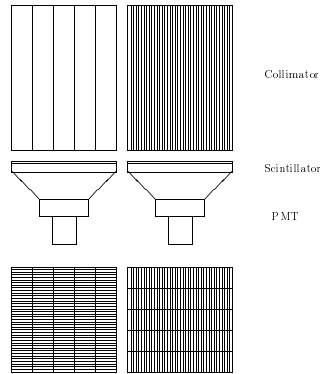,width=8cm,angle=0
}
\vspace{-5.5cm}
\caption{ A collimated scintillation telescope.Direct demodulation.}
\label{picture}
\end{figure}

  We made Monte Carlo simulations to study the imaging capability of a simple
collimated hard X-ray telescope consisting of two 700 cm$^{2}$ 
non-position-sensitive scintillators, each in viewed by just a single 
photomultiplier tube (PMT) and modulated by a slat collimator of 
$5^{o} \times 0.5^{o} $ 
and $0.5^{o} \times 5^{o} $ (FWHM) aperture separately 
(sketched in Figure 4).
The object scene (Figure 5a) in our simulations is the galactic
center region containing eight point sources which was observed by a coded
mask telescope flown on the Spacelab 2 mission (Skinner et al. 1987).
The strongest source intensity was $7.05 \times 10^{-2}$ cm$^{-2}$ s$^{-1}$, 
the ratio of the strongest source intensity to the weakest was 100:1, 
the background was taken as 0.089 counts~cm$^{-2}$ s$^{-1}$ as estimated  
from the observation. A 24 hours survey 
by the collimated telescope was simulated: pointing to various sky points
over a $6^{o} \times 6^{o}$ region around the Galactic Center separated 
by $1^{o}$ along the
galactic longitude $l$ or latitude $b$ between two successive ones, during
each pointing producing the output counts of each detector for eight 
different places of the telescope $xy$ plane rotated around the $z$ axis
by Monte Carlo simulation. Figure 5b is the cross-correlation map from the 
simulated data. A reconstruction by the direct demodulation
technique is shown in Figure 5c, which seems even better than the simulated image
of the same object scene by a 1473 cm$^{2}$ coded mask telescope with
$93 \times 99$ detector pixels for 24 hours observation (Lei et al. 1991).
 
For studying the effect of
background variation on modulation imaging, we doubled the 
background intensity when the telescope pointing to $(l,b) = (1.0^{o},-1.0^{o})$
but no recognizable change was produced in the resultant image.
In addition we assumed that starting from the pointing 
$(l,b) = (1.0^{o},-1.0^{o})$
the background intensity was linearly increasing with the passage of
time up to three times and then gradually decreasing back to the  
initial level, the whole variation lasted one hour, the corresponding 
direct reconstruction is shown in Figure 5d.
\begin{figure}
\hspace{4cm}
\psfig{figure=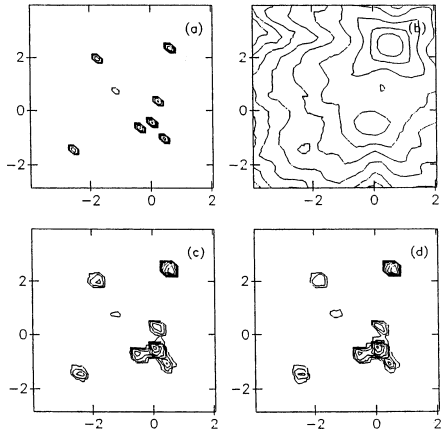,width=8cm,angle=0
}
\vspace{-5cm}
\caption{ Modulation imaging. (a) True intensity distribution.
(b) Cross-correlation map from simulated data obtained by a collimated
telescope shown in Fig.4. (c) Reconstruction by direct demodulation.
(d) Direct reconstruction for the case of variable background.}
\label{picture}
\end{figure}

  The modulation imaging has a capability to simultaneously
reconstruct both extended and discrete features in object. We
simulated scan observations of the object scene of $10^{o} \times 10^{o}$
shown in Figure 6a 
by the collimated telescope (Figure 4) with step of $1^{o}$, Figure 6b 
is the direct demodulation image from the simulated data.
\begin{figure}
\vspace{0.5cm}
\hspace{4cm}
\psfig{figure=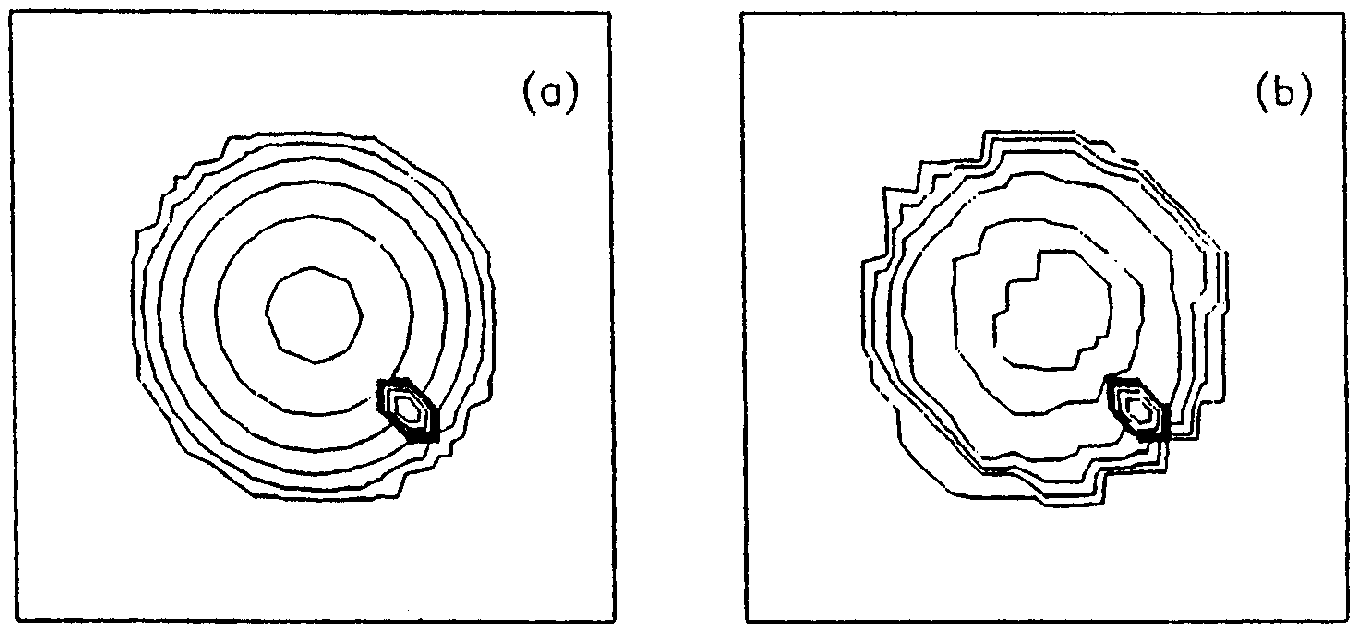,width=9cm,angle=0
}
\vspace{-9.5cm}
\caption{(a) Object scene. (b) Direct reconstruction from simulated 
data obtained by a collimated telescope shown in Figure 4.}
\label{picture}
\end{figure}

\section{Discussion}
The direct demodulation, directly solving modulation equations
and directly controlling each iterative output by physical conditions,
is a flexible technique: 
in accordance with different characters of object, instrument and
state of observation and different requirement for reconstruction
corresponding algorithms can be designed, 
and various kind of priori knowledges can be easily included in modulation
equations or constraint conditions as well.

  The modulation Equation (1), observed data $d(\mbox{\boldmath $\omega$})$ 
being an weighted
sum (integral) of an object $f(\mbox{\boldmath $x$})$ over any region with a 
weight function (integral kernel) $p(\mbox{\boldmath $\omega ,x$})$ of 
any form, can represent a measurement
process in very general case, thus the direct demodulation method can
be applied to general reconstruction problems. The CT imaging is only 
a special case of general modulation imaging:
the observed data are projections of object; in other words, 
the CT technique      
can be used only to a special case of the modulation Equation (1) where the 
integral kernel (modulation function) is a kind of $\delta$
function.

  Each iterative calculation in direct demodulation is directly based 
on the modulation equation. Obviously the direct approach can use the
information containing in the modulation equations more sufficiently 
than any indirect one through a statistical criterion or 
cross-correlation transformation, that is the main reason of the 
resolution of a direct reconstruction being able to be much better
than the intrinsic resolution (compare Figure 2i with 2e and
Figure 5c with 5b). For ideal data without error and with necessary
information of object, i.e. the modulation matrix \mbox{\boldmath $P$} being 
non-singular (observations properly arranged) and its rank $M\geq N$ 
(number of observations
large enough), solving the modulation equations will precisely reconstruct 
the object, e.g. solving $\mbox{\boldmath $Pf=d$}_{0}$ 
can exactly restore Figure 2a from
Figure 2b no mater how wide the resolution function $p(k,i_{0})$ is. 
 
  The main difficult of the direct method is the illness of solution 
causing by errors in data. Introducing physical constraints to make nonlinear
control in the process of solving modulation equations and using correlation
transformation properly can depress the effect of errors effectively 
and then make high resolution direct reconstruction realizable. 

  The physical solution resulted from solving a modulation equation system
under physical constraints is neither a solution in a strictly mathematical 
sense nor the statistical solution in a least-squares sense. Neither the
modulation equations nor the normal equations are satisfied by the physical
solution. Studying the mathematical properties 
of physical solutions is a difficult problem of nonlinear mathematics.
The quality of a direct reconstruction (sensitivity, resolution, uncertainty,
etc.) depends on the quality and quantity of information about the object 
containing in the observed data as well as the ratio of signal to noise, 
and relates to the ability of the signal modulation pattern depressing noises.
The generality of the direct demodulation technique implies that various 
kind of apparatus (detectors and modulators) constituted by different principles
can be selected for observing the same object. Until now no 
general mathematical method can guide us to designing modulators and arranging
observations, even not a clear standard to judge the quality of a 
reconstruction. How many observations are necessary for reconstruction 
depends on properties of observation system, arrangement of observations 
and characteristics of observed object
(size of reconstructed space and complexity of object), and also relates to
requirements for reconstruction. Just from resolution consideration
we may arrange observations properly and take the number $M$ of observations
sufficiently large to make the intrinsic resolution function (7)
fine enough (comparing with the achievable limit of the instrument).

  Designing detector and modulator and estimating statistical errors
and other uncertainties in a reconstruction can be aided by 
simulation calculations. Statistical uncertainties can be evaluated 
by the bootstrap method (Efron 1979; Diaconis \& Efron 1983), 
estimating other uncertainties and 
comparing different observation instruments and arrangements can be made
by Monte Carlo simulations.  
In performing direct demodulation by a computer, the computation is quite fast 
as the algorithm is constituted by only simple arithmetic operations
and conditional control statements, although a larger space of the main 
storage is often needed to store modulation coefficients and reconstruction 
results. It is feasible in many practical
cases to study the direct demodulation through simulation calculations for
Monte Carlo samples or bootstrap samples.

  The authors thank Prof. Jun Nishimura for helpful discussion.
This work was supported by the National Natural Science Foundation of China.

\end{document}